\documentclass[sigconf]{acmart}

\usepackage{multirow, makecell}
\usepackage[show]{chato-notes}
\usepackage{tablefootnote}
\usepackage{cleveref}
\usepackage{graphicx}
\usepackage{subfig}
\usepackage{pifont}
\usepackage{algorithm}
\usepackage{algpseudocode}
\usepackage{soul}
\usepackage{booktabs}
\usepackage{adjustbox}

\newcommand{\algoname}[1]{{\sc #1}\xspace}
\newcommand{\dataset}[1]{{\sf #1}\xspace}

\newcommand{\marco}{\dataset{MS-MARCO}}

\newcommand{\Rnn}{\ensuremath{R^{*}}\xspace}
\newcommand{\dnn}{\ensuremath{{d_i}^{\ast}}\xspace}
\newcommand{\faiss}{\algoname{FAISS}}
\newcommand{\Lireg}{\algoname{Reg}}
\newcommand{\Intreg}{\algoname{Reg\ensuremath{_{+int}}}}
\newcommand{\algo}{A-\ensuremath{k}NN\xspace}
\newcommand{\aknn}{A-\ensuremath{k}NN\xspace}
\newcommand{\aknnr}{A-\ensuremath{k}NN\ensuremath{_{95}}\xspace}

\newcommand{\li}{Li \emph{et al.}\xspace}

\copyrightyear{2024}
\acmYear{2024}
\setcopyright{rightsretained}
\acmConference[CIKM '24]{Proceedings of the 33rd ACM International Conference on Information and Knowledge Management}{October 21--25, 2024}{Boise, ID, USA}
\acmBooktitle{Proceedings of the 33rd ACM International Conference on Information and Knowledge Management (CIKM '24), October 21--25, 2024, Boise, ID, USA}
\acmDOI{10.1145/3627673.3679903}
\acmISBN{979-8-4007-0436-9/24/10}

\begin{document}


\title{Early Exit Strategies for Approximate $k$-NN Search in Dense Retrieval}
\author{Francesco Busolin}
\orcid{0000-0002-3235-2524}
\affiliation{%
  \institution{Ca' Foscari University}
  \city{Venice}
  \country{Italy}
}
\authornote{Corresponding author: francesco.busolin@unive.it}

\author{Claudio Lucchese}
\orcid{0000-0002-2545-0425}
\affiliation{%
  \institution{Ca' Foscari University}
  \city{Venice}
  \country{Italy}
}

\author{Franco Maria Nardini}
\orcid{0000-0003-3183-334X}
\affiliation{%
  \institution{ISTI-CNR}
  \city{Pisa}
  \country{Italy}
}

\author{Salvatore Orlando}
\orcid{0000-0002-4155-9797}
\affiliation{%
  \institution{Ca' Foscari University}
  \city{Venice}
  \country{Italy}
}

\author{Raffaele Perego}
\orcid{0000-0001-7189-4724}
\affiliation{%
  \institution{ISTI-CNR}
  \city{Pisa}
  \country{Italy}
}

\author{Salvatore Trani}
\orcid{0000-0001-6541-9409}
\affiliation{%
  \institution{ISTI-CNR}
  \city{Pisa}
  \country{Italy}
}

\renewcommand{\shortauthors}{Francesco Busolin et al.}
\begin{abstract}
Learned dense representations are a popular family of techniques for encoding queries and documents using high-dimensional embeddings, which enable retrieval by performing approximate $k$ nearest-neighbors search (\algo). A popular technique for making \algo search efficient is based on a two-level index, where the embeddings of documents are clustered offline and, at query processing, a fixed number $N$ of clusters closest to the query is visited exhaustively to compute the result set.

In this paper, we build upon state-of-the-art for early exit \algo and propose an unsupervised method based on the notion of \textit{patience}, which can reach competitive effectiveness with large efficiency gains.  
Moreover, we discuss a cascade approach where we first identify queries that find their nearest neighbor within the closest $\tau \ll N$ clusters, and then we decide how many more to visit based on our patience approach or other state-of-the-art strategies. 
Reproducible experiments employing state-of-the-art dense retrieval models and publicly available resources show that our techniques improve the \algo efficiency with up to 5$\times$ speedups while achieving negligible effectiveness losses.
All the code used is available at \url{https://github.com/francescobusolin/faiss_pEE}.

\end{abstract}

\ccsdesc[500]{Information systems}
\ccsdesc[500]{Information systems~Information retrieval}
\ccsdesc[500]{Information systems~Retrieval effectiveness}
\ccsdesc[500]{Information systems~Retrieval efficiency}

\keywords{Neural IR, Dense Retrieval, Efficiency/Effectiveness Trade-offs}


\maketitle
\section{Introduction}

The recent developments in pretrained large language models (PLM) have shown the effectiveness of learned representations for many tasks, including ad-hoc text retrieval.  In this context, one common approach relies on learned ``dense'' representations, where neural encoders are used to independently compute query and document representations in the same latent vector space. So far, two different kinds of dense representation have emerged: \emph{single}-vector representations, where queries and documents are encoded with a single embedding~\cite{ance, star,izacard2021contriever, Hofstaetter2021_tasb_dense_retrieval, Wu2022ConTextualMA, dpr,sbert}, and \emph{multi}-vector representations, where, conversely, questions and documents are represented with multiple embeddings~\cite{colbert,santhanam-etal-2022-colbertv2, plaid, colberter}. 

We focus on state-of-the-art single-vector representations.
Given a collection of pre-computed document embeddings, retrieval of relevant documents for a query embedding becomes finding the set of documents that maximizes a similarity score, e.g., inner product, or minimizes a distance, e.g., Euclidean distance. Top-$k$ retrieval thus becomes the problem of finding the $k$ closest objects to the query in the multidimensional latent vector space, i.e., the $k$ nearest neighbors ($k$-NN) points.

\vspace{1mm}
\noindent\emph{\textbf{Related Work.}}
Approximate $k$ nearest neighbors search techniques trade-off accuracy for efficiency by avoiding scanning the entire collection thus accepting some loss in result quality~\cite{TKDE2020, eccv.2010.10.1007/978-3-642-15558-1_1, chierichetti.10.1145/1265530.1265545, DBLP:conf/vldb/SingithamMR04}. 
Indexing data structures for efficient \algo search rely on quantization or hashing/binning techniques~\cite{pan2020product,andoni2008near}. They typically partition the data points into disjoint clusters and perform a two-step search.
First, the $N$ clusters whose centroids turn out to be 
closest to the query are identified, where $N$ is a static \algo hyperparameter commonly called \textit{number of probes}.
Second, these $N$ closest clusters are visited exhaustively to identify the $k$ closest data points to the query vector.
The computational cost of \algo is proportional to the number and cardinality of the clusters 
visited, while the approximation error of the results retrieved decreases by increasing the number of clusters visited.

To further reduce the computational cost of \algo, \li~\cite{li2020ann} proposed an \textit{adaptive} \aknn technique 
that dynamically chooses
the number of clusters to probe on a per-query basis. The rationale of this strategy is to reduce 
the average query latency by limiting the number of clusters visited for \textit{easy queries}, compared to the common 
\algo strategy that always probes a fixed number $N$ of clusters for all queries. 
The technique relies on a learned regression model to estimate the number of probes and may be classified as an early-exit method for \aknn, as it aims to stop the inspection of the clusters before reaching the $N^{th}$ one. Other early-exit 
 strategies for ad-hoc search have been investigated earlier \cite{EarlyBarla10, Busolin.sigIR-21,xin-etal-2020-early,xin-etal-2021-berxit,busolin.access.10311562}.
Specifically, these strategies focus on document 
re-ranking with learning-to-rank models based on additive ensembles of regression trees and passage re-ranking models based on 
pre-trained language models. All these proposals aim to improve efficiency by selectively stopping the scoring process  
for documents that are likely not included among the top-$k$ ones.
In this paper, we further elaborate the method proposed by \li~\cite{li2020ann} to introduce \textit{adaptiveness} into the \algo algorithms by proposing novel early-exit methods for retrieval systems based on state-of-the-art single-representation dense models. Reproducible experiments conducted using three state-of-the-art dense representations, i.e., STAR~\cite{star}, CONTRIEVER~\cite{izacard2021contriever}, and TAS-B~\cite{Hofstaetter2021_tasb_dense_retrieval} on the MS-MARCO dataset~\cite{nguyen2016ms} show that our techniques improve the retrieval efficiency with \textit{speedups} ranging from $4.71\times$ to $5.27\times$ over the \faiss-based~\cite{faiss} \algo baseline.
\section{Problem Statement \& Methodology}
\begin{figure}[tb]
    \centering
    \includegraphics[width=\linewidth]{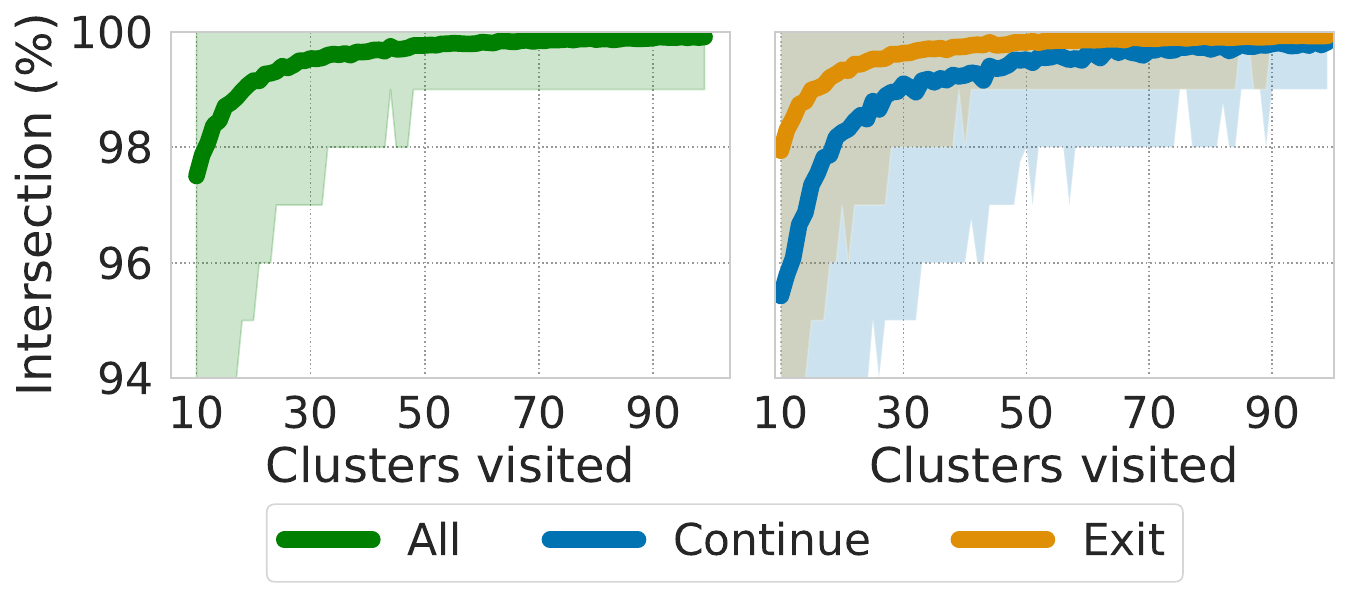}
    \caption{Average intersection size between consecutive result sets. The shaded area shows the region between the 5th and 95th percentile. The subdivision into \emph{Exit} and \emph{Continue} is obtained with $\tau = 10$.\label{fig:intersection_all}\vspace{-2mm}}
\end{figure}

\noindent\emph{\textbf{Problem Statement.} }
Given a collection of embeddings $\mathcal{D} \subset \mathbb{R}^n$ and a query  $q$ embedded in the same latent space, 
we aim at identifying the $k$ elements in $\mathcal{D}$ that are closest to $q$ according to a similarity score  $\sigma(q, d)$. In the case of dense retrieval models, the latent space is learned, and the function $\sigma(q, d)$ becomes a proxy of \emph{query-document relevance}. Therefore, the set $k\mathrm{NN}(q, \mathcal{D})$ computed using $\sigma$ likely includes the $k$ most relevant documents for $q$. However, computing the \textit{exact} result set $k\mathrm{NN}(q, \mathcal{D})$ can become computationally very expensive with large datasets composed of high dimensional vectors~\cite{Indyk98, weber98, Li2016ApproximateNN}. Thus, practical search systems commonly exploit approximate \algo algorithms, which rely on a precomputed partition of $\mathcal{D}$, and probe only the small subset of $N$ clusters whose centroids are the most similar to the query. 

Let $\mathcal{Q}$ be a set of queries and $q_i \in \mathcal{Q}$ be a generic query. We denote by $d_i$ the document of $\mathcal{D}$ the most similar to $q_i$ retrieved by an \algo algorithm for a fixed number of probes $N$, i.e., $d_i = \textit{A-}\mathrm{1NN}(q_i, \mathcal{D})$.  We also denote by $\dnn$ the actual document of $\mathcal{D}$ most similar to $q_i$ computed by a brute-force method exhaustively searching $\mathcal{D}$.
If $d_i = \dnn$, the \algo algorithm can identify the document of $\mathcal{D}$ with the highest similarity to $q_i$. 

Given the retrieval task addressed with \algo,  we can measure the \textit{recall} of a result set $\mbox{\algo}(q_i, \mathcal{D})$ of $k$ denoted by $\Rnn@k$ or $R@k$, respectively. The former is computed  against  the \textit{exact result set} $k\mbox{NN}(q_i, \mathcal{D})$, whereas
the latter by considering the set of \textit{relevant documents} for a query as identified by human assessors. 
Specifically, we denote by $\Rnn@1(q_i)$ the recall at cutoff 1 using the exact $k$NN algorithm, i.e., $\Rnn@1(q_i) = 1.0 $ if $d_i = \dnn$, and $0$ otherwise. Also, we simply denote by $\Rnn@1$ the average recall, i.e., $\sum_i \Rnn@1(q_i) / |\mathcal{Q}|$. 

A good practice in \algo algorithms is to set  $N$ as the minimum 
number of probes so that $\Rnn@1 \ge \rho$, $\rho \in [0,1]$~\cite{li2020ann}. For example, $\rho = 0.95$ means that at least $95\%$ of the queries of  $\mathcal{Q}$ returns $\dnn$ as the most similar document to $q_i$. We can surely fix a very large value of $N$ such that  $\Rnn@1 = 1.0$, but this would severely hinder query processing efficiency.
To make the \algo algorithm efficient without hampering too much effectiveness, we now discuss the baseline by \li~\cite{li2020ann} and our proposed techniques to make  \algo \textit{adaptive} for the different queries $q_i \in \mathcal{Q}$. We call this family of techniques Adaptive Approximate $k$NN methods, as they visit the
clusters \textit{ordered} by similarity to a query $q_i$, but adaptively choose, for each query, the best number of clusters to probe. The hyper-parameter $N$ of \algo is still used in the adaptive techniques as an upper bound of the number of clusters to visit by any given query.

\vspace{1mm}
\noindent\emph{\textbf{Regression Approach.}}
The regression-based approach proposed by \li~\cite{li2020ann}, which we call \Lireg, adaptively estimates the number $r(q_i)$ of probes to process a query $q_i$, where $r(q_i) \le N$.
To train and test the regression model, the query set $\mathcal{Q}$ is subdivided into train/validation/test partitions, and a golden standard $C(q_i)$ is associated with each $q_i$, where  $C(q_i) \le N$ is the minimum number of clusters to probe to find its closest vector $\dnn \in\mathcal{D}$.  If $\dnn$ cannot be found in the $N$ closest clusters, $C(q_i)=N$. 
It is worth noting that some features needed by the regression model depend on the $k$ neighbors obtained after a partial search; these features are reported in groups $(1)$, $(2)$, and $(3)$ of \autoref{tab:features}.
To this end, besides $N$, we introduce the positive parameter $\tau$, with $\tau \ll N$, that controls the number of clusters to visit by all queries to extract the features.
Generally, small values of $\tau$ would improve the overall efficiency at the expense of the effectiveness of the predictions.
Due to the imbalance of the dataset, characterized by a large fraction of queries that need to probe very few clusters, 
the learned regression model struggles to accurately predict the value of $C(q_i)$.

\vspace{1mm}
\noindent\emph{\textbf{Patience-based Approach.}} The notion of \emph{patience} has been long and widely explored for many tasks, primarily to prevent over-fitting during model training~\cite{morgan1989generalization, NIPS2000_059fdcd9, prechelt2002early, Yao2007} and for early termination of inference~\cite{NEURIPS2020_d4dd111a, zhu-2021-leebert, fpabee, zhang-etal-2022-pcee, GAO2023126690, sponner2024temporal, Sponner2023TemporalPE}. We now discuss how to make use of a \emph{patience}-based method to early-terminate \algo. During the iterative visit to the $N$ clusters sorted by similarity to query $q_i$, the $k$NN$(q_i, \mathcal{D})$ result set is progressively updated. Let $RS_h(q_i)$ be the result set of the $k$ documents obtained after visiting the first $h$ clusters. Let $\phi_{h}(q_i)$ be the size, expressed in \textit{percentage}, of the intersection between $RS_{h}(q_i)$ and the previous results set at iteration $h-1$, i.e., $\phi_{h}(q_i) = 100\cdot|RS_{h-1}(q_i) \cap RS_{h}(q_i)| / k$.

\autoref{fig:intersection_all}~(left) plots $\phi_{h}$, which is the mean $\phi_{h}(q_i)$ for all the query $q_i \in \mathcal{Q}$.  After around 30 clusters, the average $\phi_{h}$ saturates and quickly approaches $100\%$.
This saturation phenomenon suggests we can stop the visit of further clusters when the result set does not change by much for a given number $\Delta$ of iterations, e.g., when $\phi_{h}(q_i) \ge \Phi \in [90,100]$ for $\Delta$ consecutive iterations. In other words, we stop the evaluation of $q_i$ if, for $\Delta$ consecutive iterations, visiting the next cluster keeps at least $\Phi\%$ of the $k$ closest documents unchanged.
The effectiveness of this heuristic still depends on the cluster $h$ on which we stop the iterative search, which hopefully should be equal to or slightly greater than $C(q_i)$. 

\vspace{1mm}
\noindent\emph{\textbf{Classification Approach.} }
To introduce this further technique, we briefly discuss the \textit{frequency distribution} of the labels $C(q_i)$. Independently of the dense encoders adopted, we can observe that  $C(q)$ follows a \emph{power-law} distribution, where approximately $50\%$ of all the queries in $\mathcal{Q}$ need to explore just the closest cluster to find and return their nearest neighbors, i.e., for about $50\%$ of all the queries we have that $C(q_i)=1$. Moreover, about $80\%$ of the queries $q_i \in \mathcal{Q}$ only need to probe at most 10 clusters to find  $\dnn$. 

From this analysis, since most queries need only a handful of clusters to retrieve their nearest neighbor document, we could take advantage of a \textit{classifier} aimed to early identify those queries. To this end, we reuse $\tau$, to stop at the $\tau^{th}$ cluster the processing of all queries $q_i$ for which $C(q_i) \le \tau$, and proceed until $N$ for the others. To prepare the training/validation/test dataset for the classifier, we thus need to relabel the queries: a query $q_i$ is thus labeled as \emph{Exit} if $C(q_i) \leq \tau$, and as \emph{Continue} otherwise. Since the queries labeled as \emph{Exit} form the majority class, we rebalance the two classes using SMOTE~\cite{chawla2002smote}.

\autoref{fig:intersection_all}~(right) plots the average value $\phi_{h}$ as a function of the cluster visited. Indeed, we plot two curves, each relative to the queries labeled either \textit{Exit} or \textit{Continue} when $\tau = 10$. We observe how $\phi_{h}$ soon becomes saturated, approaching 100\% in both cases. Still, the curves are well separated. Also, on average, the \emph{Exit} curve shows an earlier saturation than the other and is more stable as it has less variability. This also suggests that \textit{patience}-based features can be effective for our classification task.
As such, we added these features to the ones described by \li~\cite{li2020ann} to train the classification and the regression models. The full set of features is detailed in \autoref{tab:features}.

\vspace{1mm}
\noindent\emph{\textbf{Cascade Approach}}.
If a pure classifier
can successfully detect which queries have to early exit at the $\tau^{th}$ cluster, the natural follow-up question would be about what to do with the remaining \textit{Continue} queries. A possible answer is a \textit{cascade} technique, aimed to early stop the processing of these queries possibly before the $N^{th}$ cluster. Specifically, we employ either a regression-based or a patience-based method for this second cascade stage, where the first stage is the classifier.
In this framework, note that between \textit{False Exit} and \textit{False Continue}, only the former can affect effectiveness since the classifier stops processing $q_i$ even though $\dnn$ has not yet met. The latter only reduces efficiency, as the processing of $q_i$ is not stopped and only ends when all the $N$ clusters have been explored. When we train the classifier, we are more interested in reducing the \emph{False Exits} as not to penalize effectiveness. So, in addition to using SMOTE, we also increase the training penalty of a \emph{False Exit} by weighting by a factor $w \ge 1$ the instances of the \emph{Exit} class.
This weighting strategy is particularly advantageous within a cascade approach, where a large amount of \emph{False Continues} can be early-stopped by the next cascade stage.

\begin{table}[tb]
    \caption{Features used by the learned methods. \Lireg~\cite{li2020ann} uses groups $(1)$, $(2)$, and $(3)$, while \Intreg and the Classifier employ all the features.}
    \label{tab:features}
    \footnotesize
    \centering
    \begin{tabular}{@{}c|cl@{}}
    \toprule
        Type & Feature & Description  \\
        \midrule
        Query$^{(1)}$ & 768 query values & the query vector \\  
        \midrule

    \multirow{2}{*}{Centroid$^{(2)}$} & $\sigma (q, c_h)$ & similarities of query to  \\
    & $h \in \{1.. \tau\} \cup \{10, 20,..., 100\}$ & $h$-th closest centroid \\
    \midrule
    \multirowcell{6}{Result \\after $\tau$ \\clusters$^{(3)}$} & $\sigma_{\tau} (q, d_1)$ & max doc. similarity \\
    \cmidrule{2-3}
    & $\sigma_{\tau} (q, d_k)$ & $k$-th doc. similarity \\
    \cmidrule{2-3}
    & $\sigma_{\tau} (q, d_1)$ / $\sigma_{\tau} (q, d_k)$ & ratio of max and  \\ 
    & &  $k$-th doc. similarities \\
    \cmidrule{2-3}
    & $\sigma_{\tau} (q, d_1)$ / $\sigma(q, c_1)$  & ratio of similarities of \\
    & & closest doc. and centroid \\
    
    \midrule
    \multirow{4}{*}{Stability$^{(4)}$} & $|RS_{h-1}(q_i) \cap RS_{h}(q_i)| / k$ & intersections between \\ 
    & $h \in \{2,...,\tau\}$  & consecutive results  \\
    \cmidrule{2-3}
    &  $|RS_{1}(q_i) \cap RS_{h}(q_i)| / k$ & intersections with \\ 
    & with $h \in \{2,..., \tau\}$  & first result \\
    
    \bottomrule
    \end{tabular}
\end{table}

\section{Experimental Evaluation}
\label{sec:exp}

The experiments discussed in this section aim to answer the following research questions:
\begin{itemize}
\item[\textbf{RQ1}:] Does a heuristic \textit{patience}-based method differ significantly from a learned regression-based one?
\item[\textbf{RQ2}:] Can a \textit{cascade} method improve single-stage approaches?
\end{itemize}

\vspace{1mm}
\noindent \emph{\textbf{Experimental Settings}}. We experiment on the public \marco (MAchine Reading COmprehension) Passage (ver. 1) dataset~\cite{nguyen2016ms}. 
We encode documents and queries in one $768$-dimensional dense vector using STAR~\cite{star}, CONTRIEVER~\cite{izacard2021contriever}, and TAS-B~\cite{Hofstaetter2021_tasb_dense_retrieval}, generating a single embedding for each document and query in the collection.
In this way, we build three collections of ${\sim}8$.$8$M vectors. We divide the $101$,$093$ queries into three sets to train our models. For testing, we use the official $6$,$980$ judged subset given by the \marco maintainers\footnote{\emph{dev/small}: ir-datasets.com/msmarco-passage.html\#msmarco-passage/dev}, whereas for training and validation we divide the remaining $94$,$113$ queries into training ($67\%$) and validation set ($33\%$).
We use \faiss \cite{faiss} to efficiently index and retrieve passages encoded as dense vectors.
Specifically, we build three IVF two-level indexes to partition the collections in $65$,$536$ clusters\footnote{We use the smallest power of two greater than $16 \times \sqrt{|\mathcal{D}|}=16 \times \sqrt{{\sim}8.8\mathrm{M}}$.}
each, based on the inner product between vectors.

\noindent Similar to \li~\cite{li2020ann}, we build our regression and classification models using small additive forests of $100$ trees using LightGBM~\cite{lightgbm-nips17} and use HyperOPT~\cite{hyperopt} for hyperparameter tuning with an early stopping window set to $10$.

\noindent In particular, we modified FAISS 1.7.4 and used as-is openBLAS 0.3.26 and LightGBM 4.3.0.
All our experiments were performed on a machine with $504$ GB of memory and two Intel Xeon Platinum 8276L CPU @ 2.20GHz with $56$  physical cores. To ensure accurate measurement of the execution time, we conduct each experiment $6$ times in a row, discard the initial run, and then calculate the average execution time as the average of the remaining $5$ runs.

\vspace{1mm}
\noindent\emph{\textbf{Parameter Selection}}.
Due to space considerations, we forego a comprehensive discussion of all the parameter selections and present a brief summary of the process followed to tune them. First, for all techniques, we set the parameter $k$, i.e., the size of the result set, to $100$.
We tuned parameter $\tau$ in $\{2, 5, 8, 10, 12, 15\}$ for the classifier and the regression model. The value $\tau=10$ consistently provides a good trade-off between effectiveness and efficiency. 
Then, to compare the methods, we align our proposals' effectiveness to that obtained by \Lireg~\cite{li2020ann}; we do so by choosing our parameters to minimize the scoring time and, at the same time, match the $\Rnn$@1 obtained by \Lireg. For example, considering STAR, \Lireg achieves $\Rnn@1 = 0.93$. Thus, we select our parameters to obtain the lowest execution times that show a $\Rnn@1 \ge 0.93$.
To penalize the prediction of \emph{False Exits}, we increase the instance weight $w$ of the \emph{Exit} class during the classifier's training. We tried with $1$, the default option, $3$, $5$, and $7$, and observed that $3$ permits to match \Lireg with STAR and TAS-B, while for CONTRIEVER we use $7$. We evaluated our patience-based strategy 
by trying different values of $\Delta$ and $\Phi$, i.e., $\Delta \in \{5, 7, 10, 12, 14\}$, and $\Phi \in \{90\%, 95\%, 100\%\}$. Ultimately, we chose $\Delta=7$ for STAR, $\Delta=12$ for CONTRIEVER, and $\Delta=14$ for TAS-B. Finally, we set the tolerance $\Phi = 95\%$ for all of them. 

\vspace{1mm}
\noindent\emph{\textbf{Discussion}}.
The experimental results are reported in Table~\ref{tab:results}, subdivided into three blocks, each associated with a different encoder (STAR, CONTRIEVER, and TAS-B).
The first two rows of each block report the results obtained by the two baselines, namely \algo and the Adaptive \algo using \Lireg~\cite{li2020ann}. Following the methodology of \li~\cite{li2020ann},  we 
choose the minimum value of $N$ that allows \algo to achieve a given $\Rnn@1$: we chose $\Rnn@1 =95\%$ and thus denote the strategy by \aknnr. 
For example, considering STAR, \aknnr reaches $\Rnn@1 = 95\%$ when all queries in $\mathcal{Q}$ are processed by always exploring the $N=80$ closest clusters. 
The metrics $R@100$ and $mRR@10$ refer instead to the MSMARCO relevant passages, thus estimating the "real" effectiveness of the ranked result set. Finally, $\widehat C$ and T are the per-query average number of clusters probed and 
measured execution times (in \textit{ms}), whereas Sp is the speedup obtained over the baseline \aknnr. 
 In the third row of each table block, we report the results of an additional baseline denoted as \Intreg.  \Intreg is obtained by adding to the feature set of  \Lireg the ones related to the stability of the result set, i.e., the features based on the intersection of the result sets inspired by our \textit{patience}-based heuristic technique.
As regards  \textbf{RQ1}, the \textit{patience}-based heuristic technique, denoted by Patience$_{\Delta=x}$, does not significantly differ from the \Lireg-based competitors in terms of effectiveness. However, our technique is much more efficient, with speedups ranging from 2.95$\times$ to 5.13$\times$. The shorter processing times compared to both the \Lireg-based approaches are primarily due to the fewer clusters probed, as Patience$_{\Delta=x}$ visits, on average, between $35$ and $84$ fewer clusters per query, depending on the \Lireg version and encoder used. As expected, the classifier-based approach, where the training instances are weighted, can reduce the \textit{False Exits} and is non-significantly worse than \aknnr for all three encoders.

Finally we answer \textbf{RQ2}, by observing
that a cascade method can increase, sometimes substantially, the efficiency of a non-cascade 
one. 
In particular, we compare a pure \Intreg and a pure \textit{patience}-based early exit method with the corresponding cascade ones.
For example, if we consider STAR, the relative speedups of the two cascade methods over the non-cascade ones are  $1.89\times$ and $1.39\times$. 
Analogously, with CONTRIEVER,  we observe that the relative speedups of the cascade method over a pure \Intreg and our \textit{patience}-based strategy are $2.19\times$ and $1.52\times$. Conversely, with TAS-B, we observe speedups of only $1.26\times$ and $1.10\times$. 
However, we observe that the effectiveness of cascade methods generally degrades, thus becoming significantly worse than \aknnr, with the only exception being the cascade using \Intreg on TAS-B.
In conclusion, the more effective encoders CONTRIEVER and TAS-B need a large value of $N$ to allow \algo to reach $\Rnn@1=0.95$. Thus, both can benefit greatly from adaptive methods that permit to inspect, on average, much less than $N$ clusters. For these two encoders, a pure \textit{patience}-based heuristic can achieve speedups of $4.04\times$ and $5.13\times$, getting a value of mRR@10 comparable to the one obtained by \aknnr, while the corresponding cascade method further improves the speedup at the cost of lower effectiveness. 

\begin{table}[htb]
       \caption{ Effectiveness/efficiency results. 
        Statistical significance on  mRR@10 for the EE strategy against the corresponding \aknnr is tested with a pairwise t-test with Bonferroni correction (* and ** indicate p-values of  $<0.05$ and  $<0.01$).}\label{tab:results}
    \centering
    \footnotesize
    \setlength\tabcolsep{.8\tabcolsep}%
    \begin{tabular}{@{}lllllllll@{}}
    \toprule
    &  
    & $\Rnn$@1 & \hspace{-.1cm}R@100 & \hspace{-.1cm}mRR@10 & \hspace{-.1cm}T\;(\textit{ms)} & $\widehat{C}$ & Sp \\
    \midrule
    \multirow{8}{*}{\rotatebox{90}{STAR\ \ \  \scriptsize{($N=80$)}}}
    & \aknnr          & 0.951 & 0.791 & 0.296 & 0.835 & 80.0 & 1.00\\
    & \Lireg~\cite{li2020ann}   & 0.932 & 0.769 & 0.291 & 0.737 & 69.6 &1.13\\ 
    & \Intreg         & 0.934 & 0.769 & 0.291 & 0.616& 53.9 & 1.36\\ 
    & Patience$_{\Delta=7}$          & 0.933 & 0.772 & 0.291   & 0.283 & 18.7 & 2.95 \\
    & Classifier             & 0.916 & 0.751 & 0.284** & 0.338& 25.1 & 2.47 \\
    & Classifier$_{w=3}$        & 0.930 & 0.763 & 0.289   & 0.305& 24.2 & 2.74 \\
    & \quad + \Intreg     & 0.922 & 0.756 & 0.287*  & 0.325& 23.5 & 2.57\\
    & \quad + Patience$_{\Delta=7}$   & 0.918 & 0.753 & 0.286*  & 0.203& 12.2 & 4.11 \\
    
    \midrule

    \multirow{8}{*}{ \adjustbox{minipage=30pt,rotate=90}{CONTRIEVER\\ \scriptsize{(N = 140)}}}
    & \aknnr                           & 0.950 & 0.834 & 0.316 & 1.392 & 140.0 & 1.00\\
    & \Lireg~\cite{li2020ann}          & 0.933 & 0.811 & 0.310 & 1.178 & 118.6 & 1.18\\ 
    & \Intreg                          & 0.939 & 0.817 & 0.313 & 0.969& 97.9 & 1.44\\ 
    & Patience$_{\Delta=12}$           & 0.933 & 0.812 & 0.310 & 0.345& 23.1 & 4.04\\
    & Classifier                       & 0.911 & 0.787 & 0.302** & 0.354& 25.9 & 3.93  \\
    & Classifier$_{w=7}$               & 0.939 & 0.819 & 0.311   & 0.522& 52.8 & 2.67 \\
    & \quad + \Intreg                  & 0.930 & 0.807 & 0.308*  & 0.442& 36.9 & 3.15\\
    & \quad + Patience$_{\Delta=12}$   & 0.925 & 0.801 & 0.306* & 0.227& 13.6 & 6.13 \\
    \midrule
    \multirow{8}{*}{\rotatebox{90}{TAS-B\ \ \  \scriptsize{($N=190$)}}}
    & \aknnr                      & 0.951 & 0.826 & 0.323   & 2.027 & 190.0 &1.00\\
    & \Lireg~\cite{li2020ann}        & 0.920 & 0.796 & 0.315   & 1.655& 123.1 & 1.22 \\
    & \Intreg                 & 0.928 & 0.799 & 0.316   & 1.265& 111.9 & 1.60\\
    & Patience$_{\Delta={14}}$           & 0.921 & 0.798 & 0.314   & 0.395& 28.3 & 5.13 \\
    & Classifier                   & 0.905 & 0.773 & 0.306** & 0.635& 49.2 & 3.19\\
    & Classifier$_{w=3}$           & 0.926 & 0.792 & 0.315   & 1.207& 118.5 & 1.68\\
    & \quad + \Intreg          & 0.915 & 0.780 & 0.312   & 1.007 & 81.9 & 2.01\\
    & \quad + Patience$_{\Delta={14}}$   & 0.903 & 0.772 & 0.307*  & 0.356 & 22.1 & 5.69 \\
    \bottomrule
    \end{tabular}
\end{table}

\section{Conclusions}
This work addresses adaptive \aknn in the context of dense, single-representation retrieval.
We presented a simple, fast, and adaptable heuristic method based on the concept of patience that can achieve the same effectiveness as learned regression-based approaches with lower evaluation times and higher speedups. In future work, we will explore the behavior of clustered indexes when reducing the approximation tolerance with respect to an exact, exhaustive search. We observe that as we increase the number of clusters, the margin between our \emph{patience} approach and the \Lireg-based approaches increases, and it is not clear if that is due entirely to the different encoders we considered or if it is a more general pattern.

\noindent
\textbf{Acknowledgements}.
This work was partially supported by the EU RIA project EFRA (Grant 101093026), and by the following Next Generation EU (EU-NGEU) projects: SERICS (Grant NRRP
M4C2 Inv.1.3 PE00000014), iNEST (Grant NRRP M4C2 Inv.1.5 ECS00000043), and FAIR (Grant NRRP M4C2 Inv.1.3 PE00000013). 

\clearpage
\bibliographystyle{splncs04}
\bibliography{biblio}

\begin{thebibliography}{10}
\providecommand{\url}[1]{\texttt{#1}}
\providecommand{\urlprefix}{URL }
\providecommand{\doi}[1]{https://doi.org/#1}

\bibitem{andoni2008near}
Andoni, A., Indyk, P.: Near-optimal hashing algorithms for approximate nearest neighbor in high dimensions. Communications of the ACM  \textbf{51}(1),  117--122 (2008)

\bibitem{hyperopt}
Bergstra, J., Komer, B., Eliasmith, C., Yamins, D., Cox, D.D.: Hyperopt: a python library for model selection and hyperparameter optimization. Computational Science \& Discovery  \textbf{8}(1),  014008 (2015)

\bibitem{Busolin.sigIR-21}
Busolin, F., Lucchese, C., Nardini, F.M., Orlando, S., Perego, R., Trani, S.: Learning early exit strategies for additive ranking ensembles. In: Diaz, F., Shah, C., Suel, T., Castells, P., Jones, R., Sakai, T. (eds.) {SIGIR} '21: The 44th International {ACM} {SIGIR} Conference on Research and Development in Information Retrieval, Virtual Event, Canada, July 11-15, 2021. pp. 2217--2221. {ACM} (2021). \doi{10.1145/3404835.3463088}, \url{https://doi.org/10.1145/3404835.3463088}

\bibitem{busolin.access.10311562}
Busolin, F., Lucchese, C., Nardini, F.M., Orlando, S., Perego, R., Trani, S.: Early exit strategies for learning-to-rank cascades. IEEE Access  \textbf{11},  126691--126704 (2023). \doi{10.1109/ACCESS.2023.3331088}

\bibitem{EarlyBarla10}
Cambazoglu, B.B., Zaragoza, H., Chapelle, O., Chen, J., Liao, C., Zheng, Z., Degenhardt, J.: Early exit optimizations for additive machine learned ranking systems. In: Proc. WSDM. pp. 411--420. {ACM} (2010)

\bibitem{NIPS2000_059fdcd9}
Caruana, R., Lawrence, S., Giles, C.: Overfitting in neural nets: Backpropagation, conjugate gradient, and early stopping. In: Leen, T., Dietterich, T., Tresp, V. (eds.) Advances in Neural Information Processing Systems. vol.~13. MIT Press (2000), \url{https://proceedings.neurips.cc/paper_files/paper/2000/file/059fdcd96baeb75112f09fa1dcc740cc-Paper.pdf}

\bibitem{chawla2002smote}
Chawla, N.V., Bowyer, K.W., Hall, L.O., Kegelmeyer, W.P.: Smote: synthetic minority over-sampling technique. Journal of artificial intelligence research  \textbf{16},  321--357 (2002)

\bibitem{chierichetti.10.1145/1265530.1265545}
Chierichetti, F., Panconesi, A., Raghavan, P., Sozio, M., Tiberi, A., Upfal, E.: Finding near neighbors through cluster pruning. In: Proceedings of the Twenty-Sixth ACM SIGMOD-SIGACT-SIGART Symposium on Principles of Database Systems. p. 103–112. PODS '07, Association for Computing Machinery, New York, NY, USA (2007). \doi{10.1145/1265530.1265545}, \url{https://doi.org/10.1145/1265530.1265545}

\bibitem{GAO2023126690}
Gao, X., Liu, Y., Huang, T., Hou, Z.: Pf-berxit: Early exiting for bert with parameter-efficient fine-tuning and flexible early exiting strategy. Neurocomputing  \textbf{558},  126690 (2023). \doi{https://doi.org/10.1016/j.neucom.2023.126690}, \url{https://www.sciencedirect.com/science/article/pii/S0925231223008135}

\bibitem{fpabee}
Gao, X., Zhu, W., Gao, J., Yin, C.: F-pabee: Flexible-patience-based early exiting for single-label and multi-label text classification tasks. In: ICASSP 2023 - 2023 IEEE International Conference on Acoustics, Speech and Signal Processing (ICASSP). pp.~1--5 (2023). \doi{10.1109/ICASSP49357.2023.10095864}

\bibitem{colberter}
Hofst\"{a}tter, S., Khattab, O., Althammer, S., Sertkan, M., Hanbury, A.: Introducing neural bag of whole-words with colberter: Contextualized late interactions using enhanced reduction. In: Proceedings of the 31st ACM International Conference on Information \& Knowledge Management. p. 737–747. CIKM '22, Association for Computing Machinery, New York, NY, USA (2022). \doi{10.1145/3511808.3557367}, \url{https://doi.org/10.1145/3511808.3557367}

\bibitem{Hofstaetter2021_tasb_dense_retrieval}
Hofst{\"a}tter, S., Lin, S.C., Yang, J.H., Lin, J., Hanbury, A.: {Efficiently Teaching an Effective Dense Retriever with Balanced Topic Aware Sampling}. In: Proc. of SIGIR (2021)

\bibitem{Indyk98}
Indyk, P., Motwani, R.: Approximate nearest neighbors: towards removing the curse of dimensionality. In: Proceedings of the Thirtieth Annual ACM Symposium on Theory of Computing. p. 604–613. STOC '98, Association for Computing Machinery, New York, NY, USA (1998). \doi{10.1145/276698.276876}, \url{https://doi.org/10.1145/276698.276876}

\bibitem{izacard2021contriever}
Izacard, G., Caron, M., Hosseini, L., Riedel, S., Bojanowski, P., Joulin, A., Grave, E.: Unsupervised dense information retrieval with contrastive learning (2021). \doi{10.48550/ARXIV.2112.09118}, \url{https://arxiv.org/abs/2112.09118}

\bibitem{faiss}
Johnson, J., Douze, M., J{\'e}gou, H.: Billion-scale similarity search with {GPUs}. IEEE Transactions on Big Data  \textbf{7}(3),  535--547 (2019)

\bibitem{dpr}
Karpukhin, V., Oguz, B., Min, S., Lewis, P., Wu, L., Edunov, S., Chen, D., Yih, W.t.: Dense passage retrieval for open-domain question answering. In: Proc. EMNLP. pp. 6769--6781 (2020)

\bibitem{lightgbm-nips17}
Ke, G., Meng, Q., Finley, T., Wang, T., Chen, W., Ma, W., Ye, Q., Liu, T.Y.: Lightgbm: A highly efficient gradient boosting decision tree. In: Advances in Neural Information Processing Systems. pp. 3149--3157 (2017)

\bibitem{colbert}
Khattab, O., Zaharia, M.: {ColBERT: Efficient and Effective Passage Search via Contextualized Late Interaction over BERT}. In: Proc. SIGIR. p. 39–48 (2020)

\bibitem{li2020ann}
Li, C., Zhang, M., Andersen, D.G., He, Y.: {Improving Approximate Nearest Neighbor Search through Learned Adaptive Early Termination}. In: {Proceedings of the 2020 ACM SIGMOD International Conference on Management of Data (SIGMOD)} (2020)

\bibitem{TKDE2020}
Li, W., Zhang, Y., Sun, Y., Wang, W., Li, M., Zhang, W., Lin, X.: Approximate nearest neighbor search on high dimensional data — experiments, analyses, and improvement. IEEE Transactions on Knowledge and Data Engineering  \textbf{32}(8),  1475--1488 (2020). \doi{10.1109/TKDE.2019.2909204}

\bibitem{Li2016ApproximateNN}
Li, W., Zhang, Y., Sun, Y., Wang, W., Zhang, W., Lin, X.: Approximate nearest neighbor search on high dimensional data — experiments, analyses, and improvement. IEEE Transactions on Knowledge and Data Engineering  \textbf{32},  1475--1488 (2016), \url{https://api.semanticscholar.org/CorpusID:1364239}

\bibitem{eccv.2010.10.1007/978-3-642-15558-1_1}
Mikul{\'i}k, A., Perdoch, M., Chum, O., Matas, J.: Learning a fine vocabulary. In: Daniilidis, K., Maragos, P., Paragios, N. (eds.) Computer Vision -- ECCV 2010. pp. 1--14. Springer Berlin Heidelberg, Berlin, Heidelberg (2010)

\bibitem{morgan1989generalization}
Morgan, N., Bourlard, H.: Generalization and parameter estimation in feedforward nets: Some experiments. Advances in neural information processing systems  \textbf{2} (1989)

\bibitem{nguyen2016ms}
Nguyen, T., Rosenberg, M., Song, X., Gao, J., Tiwary, S., Majumder, R., Deng, L.: Ms marco: A human generated machine reading comprehension dataset. choice  \textbf{2640}, ~660 (2016)

\bibitem{pan2020product}
Pan, Z., Wang, L., Wang, Y., Liu, Y.: Product quantization with dual codebooks for approximate nearest neighbor search. Neurocomputing  \textbf{401},  59--68 (2020)

\bibitem{prechelt2002early}
Prechelt, L.: Early stopping-but when? In: Neural Networks: Tricks of the trade, pp. 55--69. Springer (2002)

\bibitem{sbert}
Reimers, N., Gurevych, I.: {Sentence-BERT: Sentence Embeddings using Siamese BERT-Networks}. In: Proc. EMNLP. pp. 3980--3990 (2019)

\bibitem{plaid}
Santhanam, K., Khattab, O., Potts, C., Zaharia, M.: Plaid: An efficient engine for late interaction retrieval. In: Proceedings of the 31st ACM International Conference on Information \& Knowledge Management. p. 1747–1756. CIKM '22, Association for Computing Machinery, New York, NY, USA (2022). \doi{10.1145/3511808.3557325}, \url{https://doi.org/10.1145/3511808.3557325}

\bibitem{santhanam-etal-2022-colbertv2}
Santhanam, K., Khattab, O., Saad-Falcon, J., Potts, C., Zaharia, M.: {C}ol{BERT}v2: Effective and efficient retrieval via lightweight late interaction. In: Carpuat, M., de~Marneffe, M.C., Meza~Ruiz, I.V. (eds.) Proceedings of the 2022 Conference of the North American Chapter of the Association for Computational Linguistics: Human Language Technologies. pp. 3715--3734. Association for Computational Linguistics, Seattle, United States (Jul 2022). \doi{10.18653/v1/2022.naacl-main.272}, \url{https://aclanthology.org/2022.naacl-main.272}

\bibitem{DBLP:conf/vldb/SingithamMR04}
Singitham, P.K.C., Mahabhashyam, M.S., Raghavan, P.: Efficiency-quality tradeoffs for vector score aggregation. In: Nascimento, M.A., {\"{O}}zsu, M.T., Kossmann, D., Miller, R.J., Blakeley, J.A., Schiefer, K.B. (eds.) (e)Proceedings of the Thirtieth International Conference on Very Large Data Bases, {VLDB} 2004, Toronto, Canada, August 31 - September 3 2004. pp. 624--635. Morgan Kaufmann (2004). \doi{10.1016/B978-012088469-8.50056-5}, \url{http://www.vldb.org/conf/2004/RS17P1.PDF}

\bibitem{Sponner2023TemporalPE}
Sponner, M., Ott, J., Servadei, L., Waschneck, B., Wille, R., Kumar, A.: Temporal patience: Efficient adaptive deep learning for embedded radar data processing. ArXiv  \textbf{abs/2309.05686} (2023), \url{https://api.semanticscholar.org/CorpusID:261696875}

\bibitem{sponner2024temporal}
Sponner, M., Servadei, L., Waschneck, B., Wille, R., Kumar, A.: Temporal decisions: Leveraging temporal correlation for efficient decisions in early exit neural networks. arXiv preprint arXiv:2403.07958  (2024)

\bibitem{weber98}
Weber, R., Schek, H.J., Blott, S.: A quantitative analysis and performance study for similarity-search methods in high-dimensional spaces. In: Proceedings of the 24rd International Conference on Very Large Data Bases. p. 194–205. VLDB '98, Morgan Kaufmann Publishers Inc., San Francisco, CA, USA (1998)

\bibitem{Wu2022ConTextualMA}
Wu, X., Ma, G., Lin, M., Lin, Z., Wang, Z., Hu, S.: Contextual masked auto-encoder for dense passage retrieval. In: AAAI Conference on Artificial Intelligence (2022), \url{https://api.semanticscholar.org/CorpusID:251594591}

\bibitem{xin-etal-2020-early}
Xin, J., Nogueira, R., Yu, Y., Lin, J.: Early exiting {BERT} for efficient document ranking. In: Proceedings of SustaiNLP: Workshop on Simple and Efficient Natural Language Processing. pp. 83--88. Association for Computational Linguistics, Online (Nov 2020). \doi{10.18653/v1/2020.sustainlp-1.11}, \url{https://aclanthology.org/2020.sustainlp-1.11}

\bibitem{xin-etal-2021-berxit}
Xin, J., Tang, R., Yu, Y., Lin, J.: {BER}xi{T}: Early exiting for {BERT} with better fine-tuning and extension to regression. In: Proceedings of the 16th Conference of the European Chapter of the Association for Computational Linguistics: Main Volume. pp. 91--104. Association for Computational Linguistics, Online (Apr 2021). \doi{10.18653/v1/2021.eacl-main.8}, \url{https://aclanthology.org/2021.eacl-main.8}

\bibitem{ance}
Xiong, L., Xiong, C., Li, Y., Tang, K.F., Liu, J., Bennett, P., Ahmed, J., Overwijk, A.: Approximate nearest neighbor negative contrastive learning for dense text retrieval. In: Proc. ICLR (2021)

\bibitem{Yao2007}
Yao, Y., Rosasco, L., Caponnetto, A.: On early stopping in gradient descent learning. Constructive Approximation  \textbf{26}(2),  289--315 (Aug 2007). \doi{10.1007/s00365-006-0663-2}, \url{https://doi.org/10.1007/s00365-006-0663-2}

\bibitem{star}
Zhan, J., Mao, J., Liu, Y., Guo, J., Zhang, M., Ma, S.: Optimizing dense retrieval model training with hard negatives. In: Proc. SIGIR. p. 1503–1512 (2021)

\bibitem{zhang-etal-2022-pcee}
Zhang, Z., Zhu, W., Zhang, J., Wang, P., Jin, R., Chung, T.S.: {PCEE}-{BERT}: Accelerating {BERT} inference via patient and confident early exiting. In: Carpuat, M., de~Marneffe, M.C., Meza~Ruiz, I.V. (eds.) Findings of the Association for Computational Linguistics: NAACL 2022. pp. 327--338. Association for Computational Linguistics, Seattle, United States (Jul 2022). \doi{10.18653/v1/2022.findings-naacl.25}, \url{https://aclanthology.org/2022.findings-naacl.25}

\bibitem{NEURIPS2020_d4dd111a}
Zhou, W., Xu, C., Ge, T., McAuley, J., Xu, K., Wei, F.: Bert loses patience: Fast and robust inference with early exit. In: Larochelle, H., Ranzato, M., Hadsell, R., Balcan, M., Lin, H. (eds.) Advances in Neural Information Processing Systems. vol.~33, pp. 18330--18341. Curran Associates, Inc. (2020), \url{https://proceedings.neurips.cc/paper_files/paper/2020/file/d4dd111a4fd973394238aca5c05bebe3-Paper.pdf}

\bibitem{zhu-2021-leebert}
Zhu, W.: {L}ee{BERT}: Learned early exit for {BERT} with cross-level optimization. In: Zong, C., Xia, F., Li, W., Navigli, R. (eds.) Proceedings of the 59th Annual Meeting of the Association for Computational Linguistics and the 11th International Joint Conference on Natural Language Processing (Volume 1: Long Papers). pp. 2968--2980. Association for Computational Linguistics, Online (Aug 2021). \doi{10.18653/v1/2021.acl-long.231}, \url{https://aclanthology.org/2021.acl-long.231}

\end{thebibliography}
\end{document}